# Broad-band Mid-infrared Laser Generation via Cascading Deceleration in Plasma Channels

Tian-Liang Zhang[1], Yun-Xiao He[1,2], Shuang Liu[1], Jiu-Cheng Chen[1], Jian-Fei Hua[1*], and Wei Lu[1,3†]

1. Department of Engineering Physics, Tsinghua University, Beijing 100084, China
2. National Key Laboratory of Plasma Physics, Laser Fusion Research Center, China Academy of Engineering Physics, Mianyang, Sichuan 621900, China
3. Institute of High Energy Physics, Chinese Academy of Sciences, Beijing 100049, China

Corresponding authors: *jfhua@tsinghua.edu.cn, †weilu@ihep.ac.cn

## ABSTRACT

Plasma-based mid-infrared (MIR) laser generation has garnered significant interest owing to its advantage of high output power, continuous wavelength tunability, and ultrashort pulse durations. However, existing methodologies predominantly depend on high-intensity inputs at the hertz frequency level, with spectral energy concentrated near the central frequency, rendering them unsuitable for spectroscopic applications. This paper proposes and demonstrates a cascaded deceleration scheme that enables the generation of broadband MIR lasers with low energy inputs compatible with high-repetition-rate laser systems. By confining the input laser within a plasma channel, this approach preserves the laser intensity, which not only sustains the decelerating field strength but also enables the cumulative effect of deceleration across multiple distinct bubbles. Numerical simulations demonstrate that more than 30% of the 23 mJ input energy is converted into a broadband MIR output spanning wavelength from 0.58 to 6.86 μm, achieving peak powers on the order of gigawatts. The output exhibits unique time-frequency characteristics, defined by spectral sub-bands organized in a temporal sequence, wherein each sub-band comprises few-cycle pulses. Parametric analyses reveal that the spectral bandwidth broadens with increasing laser intensity, provided that the plasma density being adequate to ensure a sufficiently short deceleration length. This approach provides a practical, efficient route to broadband ultra-intense mid-infrared sources, promising for applications in Fourier transform spectroscopy and laser-induced electron diffraction.

## Introduction

Ultrashort, intense mid-infrared (MIR) lasers have emerged as pivotal scientific tools, uniquely combining wavelengths from 2 to 20 micrometers with femtosecond-scale temporal resolution. The spectral range encompasses the so-called "molecular fingerprint region" [1], enabling the direct investigation of fundamental vibrational modes and supporting applications across biological, medical and environmental science, such as infrared spectroscopy [2-4] and optical coherence tomography [5]. Moreover, the increased ponderomotive potential, which scales with the square of the wavelength, enhances the charge yield in laser-driven particle accelerations [6,7] and facilitates the production of attosecond X-rays via high-harmonic generation [8,9]. Its ultrafast temporal resolution, meanwhile, enables advanced techniques such as time-resolved imaging [10], laser micromachining [11] and transient states analyzation in chemical reactions [12,13].

MIR lasers can be generated through various optical techniques, such as $CO_2$ lasers [14], fiber lasers [15], as well as nonlinear processes including optical parametric amplification, difference frequency generation, and filamentation [16-19]. In recent years, plasma-based techniques for generating MIR lasers have attracted considerable interest owing to their notable advantages, including unlimited output power, continuous wavelength tunability, and the capability to produce ultrashort pulses [20-26]. Plasma based MIR generation is achieved through the redshift of a laser pulse, caused by energy depletion during wakefield excitation [27,28]. Given the absence of a damage threshold, plasma enables output pulses to reach energies of tens of millijoules and terawatt peak powers with joule-level input [21]. Moreover, due to the highly controllable laser-plasma interaction strength, the output wavelength can be continuously tuned across the mid-infrared to THz range by adjusting the plasma density and input parameters [22-24]. Additionally, the duration of the redshifted laser is constrained by the wakefield structure, enabling the generation of pulses with a single cycle or shorter [21,23]. These distinctive properties make plasma-based MIR sources well-suited for demanding applications such as strong-field physics [29] and ultrafast spectroscopy [12,30], underscoring their considerable scientific importance.

Current researches on plasma-based MIR sources have predominantly focused on joule-level input lasers [21-24]. However, these hundred-terawatt-class lasers operate at repetition rates of only a few hertz, rendering them incompatible for applications requiring kilohertz or higher frequencies [10,29,31]. To achieve kilohertz repetition rates, it is necessary to reduce the pulse energy to the order of tens of millijoules, owing to limitations in average laser power [32-34]. Yet, at these lower energies, the resulting weak deceleration fields are insufficient to generate radiation that broadly covers molecular vibrational energy levels. Although employing a secondary high-intensity laser could create stronger deceleration fields, this approach demands excessively high total energy and imposes impractical synchronization requirements, including precise temporal alignment corresponding to the plasma bubble length [26]. Furthermore, while significant progress has been made in broadly tunable output across the MIR range, existing plasma-based MIR radiation is spectrally concentrated near the central wavelength. This narrow bandwidth poses a particular challenge for applications like Fourier transform infrared (FTIR) spectroscopy, where the ability to probe several molecular transitions concurrently would greatly enhance analytical capability [35-37]. Thus, a critical unresolved challenge remains in the development of a novel deceleration scheme capable of efficiently producing ultrabroadband MIR radiation directly from low-energy driver lasers to unlock practical high-repetition-rate, high-resolution detection applications.

To address this challenge, this study proposes a cascaded deceleration scheme generating broadband MIR lasers with much lower energy requirement, which is compatible with high-reptition-rate laser systems. This method utilizes a plasma channel to confine the decelerated laser, maintaining sufficient intensity to displace electrons at the tail of a plasma bubble, thereby allowing the laser to enter a subsequent bubble for further deceleration without being accelerated itself. This approach not only overcomes the limitation of weak deceleration fields associated with mJ-level input lasers but also enables the generation of broadband MIR radiation spanning several micrometers in wavelength. Simulation results validate the scheme, demonstrating that over 30% of the input energy is converted into the output laser, with a spectral range extending from 0.58 μm to 6.86 μm. Moreover, the output demonstrates distinct time-frequency properties, wherein laser components at different wavelengths consist of few-cycle pulses that are temporally ordered in sequence. Parametric studies reveal that the spectral coverage of MIR laser increases with laser intensity, provided that the plasma density is such that the dispersion length remains shorter than the deceleration length. Future improvements in laser technology are expected to generate ultra-broadband mid-infrared light with spectral coverage of entire MIR region. Our work provides a practical and efficient route to high-repetition-rate MIR sources, which are urgently needed for applications such as fourier-transform infrared spectrocopy, laser-induced electron diffraction, and multicolor waveform synthesis [35-41].

## Results

We propose a cascaded laser deceleration method to achieve plasma-based, multiple-octave-spanning broadband MIR laser generation with significantly reduced energy input requirements (Fig. 1). The realization of this method depends on the confinement of the laser pulse within a plasma channel. The confinement preserves the laser's energy, enabling it to expel electrons situated at the rear of the plasma bubble, which leads to the merging of successive bubbles. Consequently, the decelerated laser can penetrate subsequent bubbles without undergoing acceleration, allowing the deceleration effects from multiple stages to accumulate. While the deceleration achieved in an individual stage is relatively limited, the cumulative impact of this cascading process can fully decelerate the laser. The laser outputs from each deceleration stage simultaneously present in a unified beam, ultimately generating ultra-broadband MIR characterized by a temporally ordered frequency-time structure.

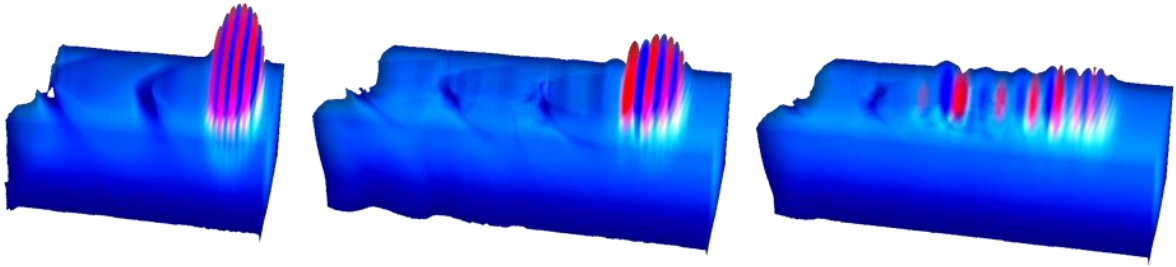

Figure 1 Schematic diagram of cascade deceleration method

**Limitations of laser energy on output wavelength in single-stage deceleration**

The output wavelength of single stage plasma-based MIR generation depends on the degree of modulation experienced by the input laser within the plasma, which can be analyzed starting from the principle of photon deceleration. The process of photon deceleration arises as a consequence of laser self-modulation, a phenomenon induced by the refractive index gradient. The deceleration speed $v_{rs}$, defined as the relative change in wavenumber $k$ per unit transmission distance, is associated with the refractive index $\eta$ as:

$$v_{rs} = -\frac{1}{k}\frac{\partial k}{\partial (ct)} = -\frac{1}{\eta^2}\frac{\partial \eta}{\partial \xi} \tag{1}$$

where $c$ is the light speed, $z$ and $t$ is the spatial and temporal coordinates, and $\xi = z - ct$. According to one-dimensional nonlinear wakefield theory [42], the refractive index distribution within the laser-induced wakefield can be derived from the electrostatic potential $\phi$ and the normalized vector potential $a$ of the laser, as shown below:

$$\eta = \frac{n}{n_0 \gamma} \cong 1 - \frac{k_p^2}{2k^2}\frac{1}{1+\phi} \tag{2}$$

$$\frac{\partial^2 \phi}{\partial \xi^2} = \frac{k_p^2}{2}\left[\frac{1+a^2}{(1+\phi)^2} - 1\right] \quad (3)$$

$k_p$ is the plasma wavenumber. Eq. (3) gives periodic solutions for $\phi$ in laser-free regions, indicating an alternating wakefield of deceleration and acceleration. Within this wakefield, the decelerated portion of the laser redshifts and shifts backward. It eventually moves into an accelerating region after a certain propagation distance, which marks the completion of the deceleration process.

Based on these equations, figure 2a illustrates the distributions of deceleration speeds, while figure 2b presents the output wavelength and required propagation distance as functions of the laser's normalized vector potential, assuming a fixed deceleration field according to fig. 2a. These two figures illustrate that a strong decelerating field and a long output wavelength necessitates a high normalized vector potential. However, at low laser energies, the attainable normalized vector potential is considerably constrained. For example, in lasers operating at kilohertz repetition rates, the pulse energy is limited to around 50 millijoules due to the average power constraints of existing laser technologies [32]. As a result, the normalized vector potential only reaches 2.4 even if compressed to few-cycle duration of 10 fs and focused to wavelength-order size of 5 μm. This limits the maximum achievable wavelength through single-stage deceleration to about 3.3 μm, shorter than characteristic peak wavelengths of common gases such as $CO_2$, NO, and CO [36]. Moreover, the excessively small focal spot yields a Rayleigh length of a hundred micrometers, which is considerably less than the deceleration distance of one millimeter required in Figure 2b. Considering the self-focusing effect is suppressed by the short pulse duration and moderate peak power, severe diffraction prevents the laser from achieving single-stage deceleration, resulting in an output wavelength even shorter than the predicted value of 3.3 μm. These limitations, arising from the inherently low energy of input lasers, render existing deceleration methods insufficient and underscore the necessity for the development of a new deceleration approach.

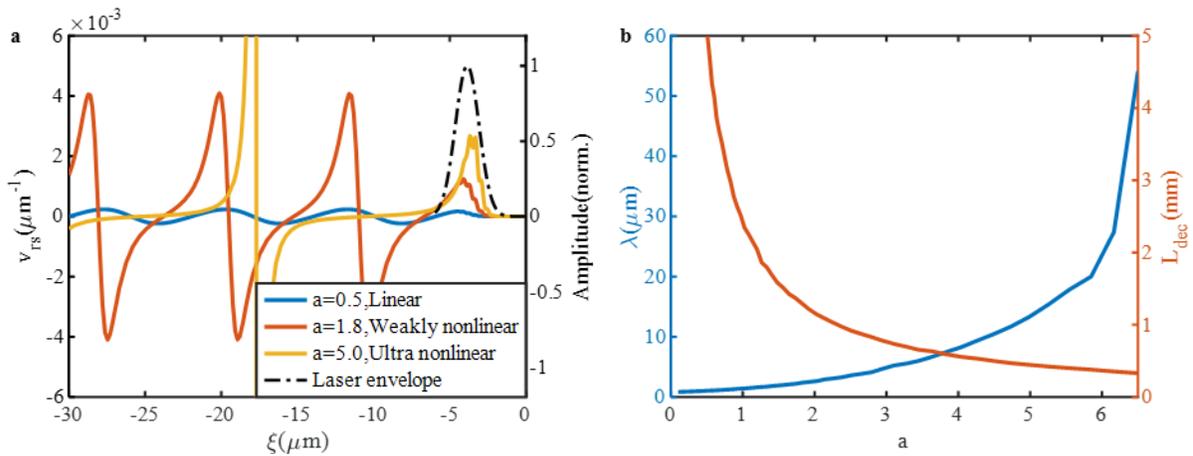

Figure 2 Dependence of deceleration dynamics on laser intensity. **a** Deceleration fields for different normalized vector potentials. **b** Required deceleration distance and final wavelength of the laser as functions of a.

**Cascading deceleration scheme**

As demonstrated in the preceding analysis, the limitations imposed on deceleration process by input laser energy are mainly in two aspects: limited propagation distance and insufficient deceleration field strength. These can be addressed by employing plasma channel confinement and accumulate the deceleration effects from multiple stages.

The limitation on transmission distance imposed by laser diffraction can be effectively overcome through plasma channel confinement. This method relies on a transverse refractive index gradient to focus the laser, a mechanism independent of the laser's power and wavelength. As a result, all wavelength components remain confined even at reduced energy levels, thereby enabling extended propagation.

The limits on output wavelength can be overcome by accumulate the deceleration effects, a process that can be described as cascading deceleration. Cascading deceleration requires the laser to penetrate the subsequent bubble without acceleration from the trailing edge of the preceding bubble, which is satisfied when the decelerated laser have enough intensity. According to the photon number conservation principle, which implies $a \propto \lambda^{0.5}$ in the case of no energy leaks, the normalized vector potential continuously rises as the laser propagates. As a result, the refractive index approaches unity with a near-zero gradient, as indicated by Eq. 3. Physically, this corresponds to the laser displacing electrons in the tail of the wakefield, resulting in the merging of two adjacent plasma bubbles. Consequently, the laser can proceed into the subsequent deceleration stage without experiencing acceleration, thereby facilitating the realization of cascading deceleration.

A simulation was performed using OSIRIS [44] to demonstrate the feasibility of cascaded deceleration (Fig. 3). The simulation employed an 800 nm laser pulse with 23 mJ energy and 10 fs duration, focused to a 5.3 μm spot radius, yielding a normalized vector potential of 1.8. The plasma was characterized by a parabolic channel density profile with a central density of $1.9 \times 10^{19}$ cm$^{-3}$ and a matching radius of 5.1 μm, slightly smaller than the laser radius to account for energy reduction during deceleration. This plasma channel can be created by ionizing the gas with a Gaussian beam focused to an 11 μm spot size, requiring less than 0.5 mJ energy. Thus, the total energy requirement remains approximately unchanged.

The simulation results demonstrate the merging of bubbles and the realization of cascade deceleration. Upon entering the plasma, the laser induces a weakly nonlinear wakefield and experiences deceleration. After propagating 300 μm, the laser reached the trailing edge of the first bubble, with its central wavelength shifting to 1.4 μm and the maximum wavelength reaching 2.2 μm (Figures 3b, e), consistent with theoretical predictions for single-stage deceleration in Fig 2b. At this point, the laser's normalized vector potential significantly exceeds unity, expelling electrons at the bubble's trailing edge and merging the first two bubbles into a unified structure. The decelerated portion of the laser, which has a lower group velocity than the leading part, moves backward into the second bubble and undergoes further deceleration with a laser wavelength extends to approximately 3 micrometers, accompanied by a normalized vector potential of 3 (Fig. 3c, f). The time-frequency distribution (Figure 3f) confirms continuous frequency reduction, with no indication of acceleration by the field at the tail of the first bubble.

The cascaded deceleration process not only accumulates wavelength variations, but also utilizes the spatially non-uniform deceleration field to produce an ultrabroadband spectrum with a unique temporal structure. In each deceleration stage, the leading edge of the pulse remains undecelerated, while the central portion is redshifted and serves as input of next stage. As a result, the original spectral components are preserved, while new wavelength components are continuously generated behind the existing ones. This leads to the formation of an ultrabroadband output with unique temporal–spectral structure, in which different wavelengths are arranged sequentially in time from the front to the rear of the pulse. Moreover, each newly generated spectral band remains confined within its corresponding bubble (Fig. 3b, c, g, h), ensuring that each spectral band of the output laser consists of few-cycle pulses.

The simulated evolution of the temporal-spectral distribution confirms this physical interpretation. At propagation distances of 480, 600, and 690 μm (Fig. 3d-f, i-j), new spectral components emerge sequentially at wavelengths of 3.0, 4.2, and 5.6 μm, respectively, while the original frequency components persist. In the temporal-spectral profiles, these newly generated bands consistently appear behind the preceding laser field, appearing as a distinct oblique line that visually demonstrates the unique wavelength-time correlation.

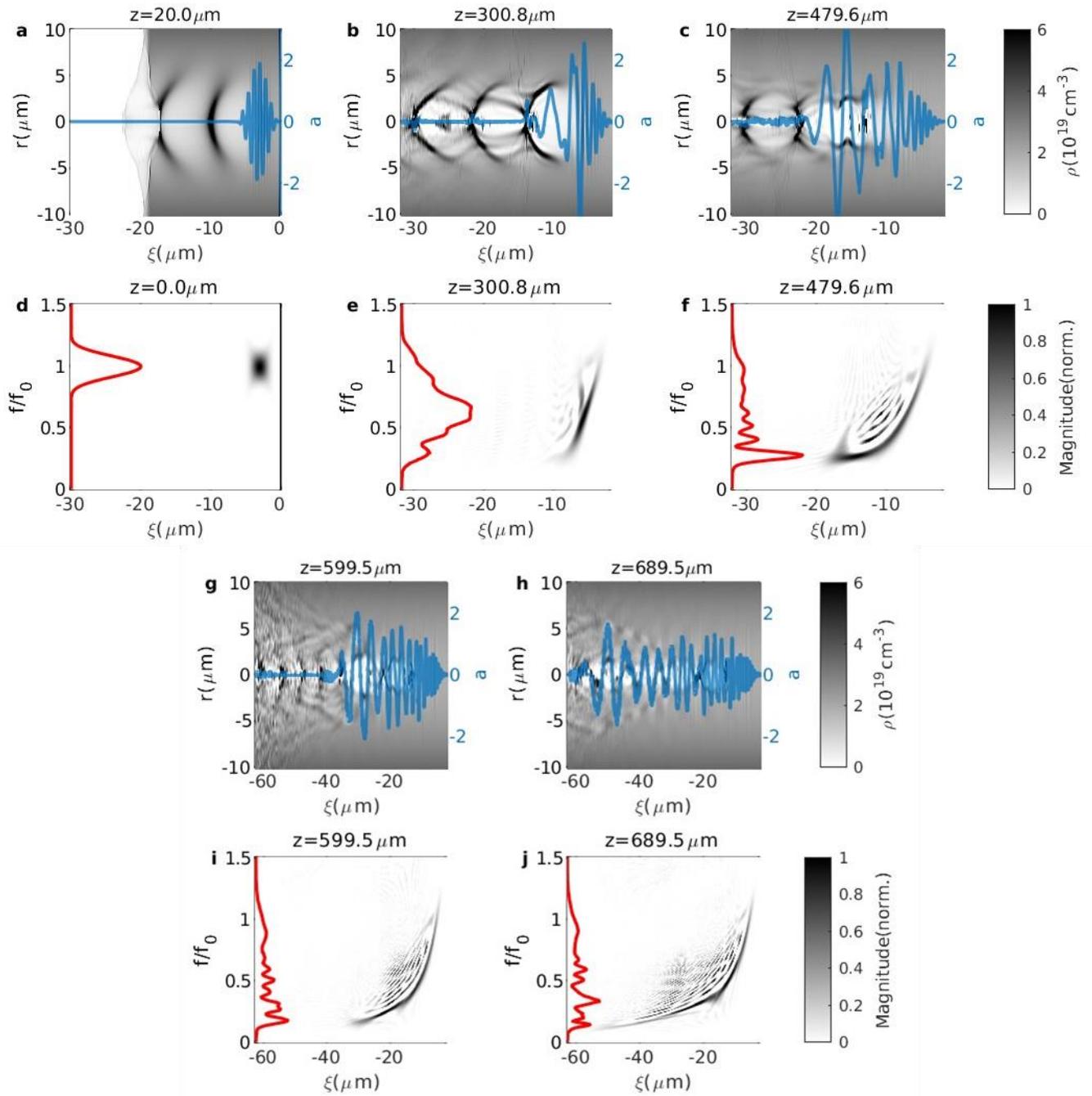

Figure 3 Simulation results of the cascading deceleration scheme. **a-c, g-h** Plasma density and the normalized vector potential of the laser at **a** 20 μm **b** 301 μm **c** 480 μm **g** 600 μm **h** 690 μm. **d-f, i-j** Corresponding time-frequency

distribution of the laser at each position.

As mentioned above, the confinement of the plasma channel is a necessary factor for achieving cascading deceleration. Plasma channels prevent laser beams from diverging caused by diffraction, thereby fulfilling the prerequisite for the merging of bubbles. The necessity of plasma channel confinement is evident from the comparative evolution of laser parameters with and without channels, as illustrated in Figures 4a and 4b. Under channel confinement, the laser radius remains approximately constant around 5 μm, while the normalized vector potential demonstrates a stepwise increase. This observed rise in the normalized vector potential, resulting from the elongation of the wavelength during deceleration process, facilitates the occurrence of the cascade process.

In contrast, in the absence of channel confinement, the laser beam radius expands rapidly as a result of diffraction, leading to a continuous reduction in its normalized vector potential. Consequently, the wakefield transitions into linear regime after 345 μm, at which point the deceleration process terminates. The laser spectrum extends only up to a maximum wavelength of 1.78 μm (Fig.4c), signifying that the first stage of deceleration remains incomplete. These findings are consistent with theoretical predictions and underscore the critical role of laser channel confinement in cascade deceleration.

Although the channel effectively confines the laser, simulation results indicate that the normalized vector potential begins to decrease after 450 μm. This is a consequence of dispersion. Approximate the bubble as a channel characterized by a central density of zero, the group velocity and dispersion length of a matched laser is described by the following equations [45]:

$$\begin{cases} \beta_g^2 = \frac{v_g^2}{c^2} = 1 - \frac{4c^2}{\omega^2 r_0^2} \\ Z_D = \frac{kL^2}{2(1-\beta_g^2)} = Z_R \left(\frac{\pi L}{\lambda}\right)^2 \end{cases} \quad (4)$$

$Z_R$ is the Rayleigh length. Since the dispersion length scales inversely with the square of the wavelength, the continuous redshifting of the laser during propagation accelerates the temporal broadening of the pulse. As a result, beyond a certain propagation distance, dispersion has a greater effect than redshift on the normalized vector potential, causing it to decrease.

The deceleration process ceases when the laser's normalized vector potential decreases to a level insufficient to expel electrons located at the tail of the bubble. In the simulation presented, this cessation occurs at 690 micrometers. At this point, the laser has undergone four stages of deceleration, with the longest wavelength reaching 6.86 micrometers. The output pulse exhibited a total energy of 8.45 mJ and a peak power of approximately 140 GW, accounting for 36.7% of the initial input energy. As shown in Figure 4a, the spectrum at the −20 dB level spans from 0.58 μm to 6.86 μm. Within this broad range, the energy is distributed across specific bands as follows: over 4.7 μm: 0.75 mJ; 2.9–4.7 μm: 0.96 mJ; 1.75–2.9 μm: 2.15 mJ. Furthermore, Figure 4b displays the electric field distribution after 690 μm of propagation, together with the temporal positions of each frequency component. The leading part of the pulse lies in the near-infrared region around 0.8 μm, followed by three mid-infrared segments with distinct central wavelengths and few-cycle structures.

This simulation result validates the feasibility of the cascaded deceleration scheme. The cascaded deceleration scheme not only addresses the limitations of insufficient deceleration in a single stage, but its unique physical process also results in a broadband laser characterized by temporally ordered few-cycle sub-spectral bands.

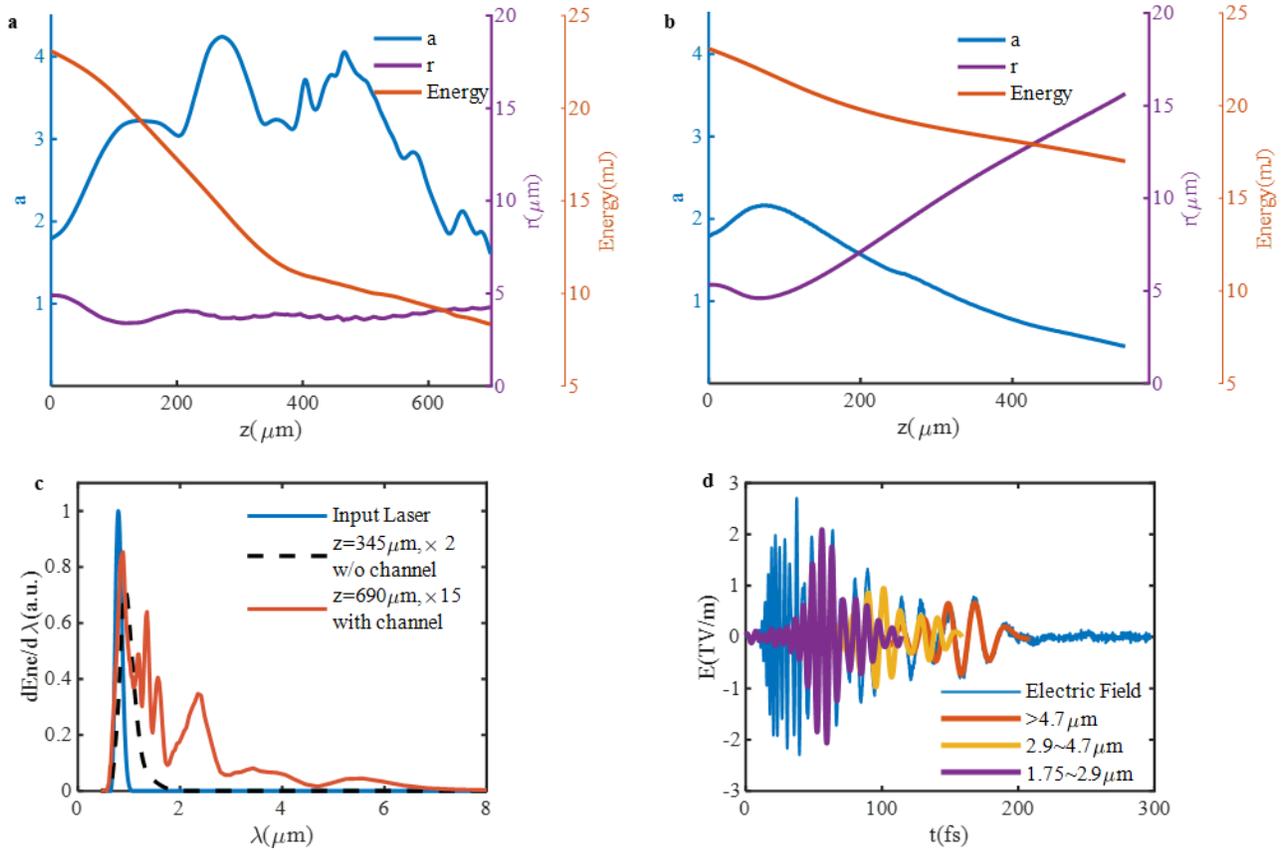

Figure 4 Laser parameters evolution during laser cascade deceleration and output spectrum. **a, b** Evolution of normalized vector potential, radius and energy with and without plasma channel, respectively. **c** Input and output laser spectrum. **d** On axis electric field of the laser pulse at z = 690 μm with plasma channel confinement, along with the contributions from individual spectral sub-bands.

**Parametric study**

To optimize the mid-infrared light output generated through cascaded deceleration, this study examines two critical parameters associated with the decelerating field strength: the on-axis plasma density $\rho_{cen}$ and the laser normalized vector potential $a$. Their effects on the spectral range, energy efficiency and the required propagation length for the deceleration process are shown in Figure 5.

Firstly, Fig. 5 shows that the relationship among the deceleration length $L_{dece}$, $\rho_{cen}$ and $a$ can be described by $L_{dece} \propto \rho_{cen}^{-1.5} a^{-1}$, which corresponds with the findings obtained from single-stage deceleration analyses [43]. This suggests that the cascaded deceleration process can be interpreted as the accumulation of single-stage decelerations.

Secondly, the variation in energy efficiency corresponds to the intrinsic physical characteristics of the deceleration process, during which the laser loses energy to elongate its wavelength. A broader spectrum indicates greater energy loss by the laser, leading to a reduced proportion of output MIR energy to the input energy.

Thirdly, the maximum wavelength component of the output MIR laser is approximately proportional to the square root on the initial normalized potential ($a_0$) in the nonlinear regime and stabilizes at an approximately constant value beyond a certain plasma density threshold. As mentioned earlier, the termination of the cascaded deceleration is attributed to the intensity reduction resulting from dispersion. A higher $a_0$ can increase the wavelength change at each stage, consequently postponing the onset of the decrease in the normalized vector potential. Conversely, although increased density also leads to greater deceleration strength, the variation in wavelength at each stage remains relatively constant because of the reduced size of the deceleration region, thereby constraining the potential for extending the output wavelength. Consequently, a stronger $a_0$ facilitates the generation of a broader spectral bandwidth MIR, while the plasma density must be optimized so that the deceleration process completes before severe dispersion occurs.

In summary, the normalized vector potential of the laser must be maximized while remaining within the constraints imposed by the available laser energy. Furthermore, the selection of plasma density should be such that the deceleration length exceeds the characteristic dispersion propagation distance. Additionally, the plasma length should be approximately matched to the required deceleration length.

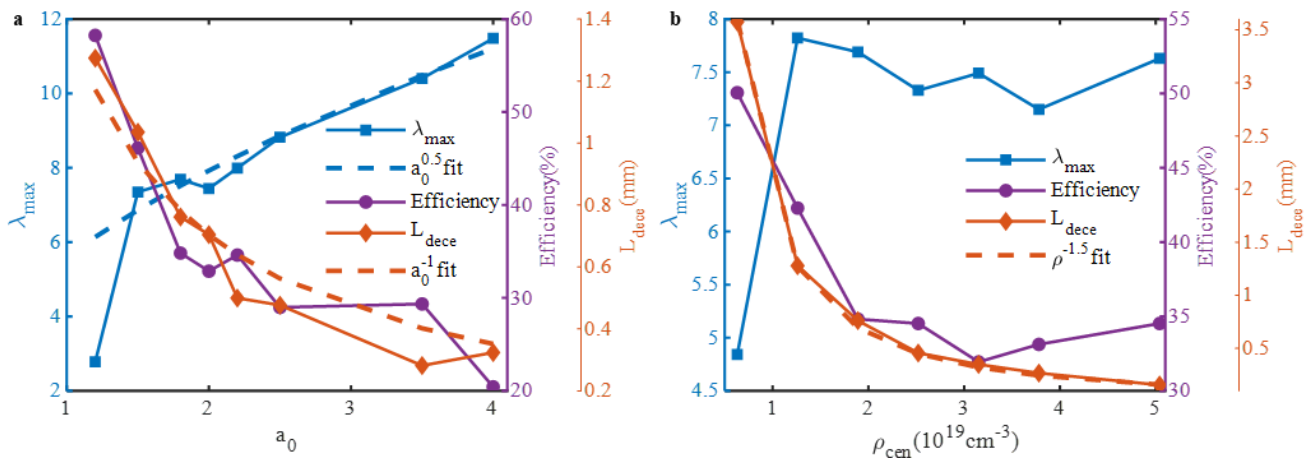

Figure 5 Relationship between the longest wavelength of spectrum $\lambda_{max}$ (-20dB), MIR conversion efficiency, deceleration length $L_{dece}$ and initial normalized vector potential $a_0$ (a) as well as plasma central density $\rho_{cen}$ (b). The $\rho_{cen}$ in panel a is $1.9 \times 10^{18} cm^{-3}$, and the $a_0$ in panel b is 1.8.

## Discussion

In conclusion, this paper proposes and demonstrates a plasma-based cascading deceleration scheme for generating ultrabroadband MIR lasers. This method utilizes plasma channels to confine the laser, enabling sufficient intensity to induce bubble coalescence. As a result, the laser undergoes deceleration within each bubble, with the deceleration effects accumulating to ultimately produce ultra-broadband mid-infrared light covering a wavelength range of several micrometers. This scheme not only overcomes the energy requirement limitations of single-stage deceleration in previous studies, but also achieves a spectrum capable for simultaneous multispecies detection. Additionally, the output MIR laser exhibits a temporal sequencing of its sub-bands, wherein shorter wavelengths precede longer wavelengths. Each sub-band is composed of few-cycle pulses, which can be filtered to produce multiple simultaneous beams at distinct wavelengths. These findings establish the potential of this scheme as a high-repetition-rate, single-source, multi-beam ultrafast intense light source.

The value of the cascaded deceleration method proposed in this paper is not limited to low input energy scenarios on the order of tens of millijoules. As indicated in the parametric study, higher energy can significantly broaden the spectral coverage of mid-infrared light, while maintaining high conversion efficiency and laser intensity, further highlighting its advantages. According to the fitting in Figure 5a, the maximum wavelength of the output laser is proportional to the square root of the initial normalized vector potential. For example, a maximum wavelength of 16 μm is possible to be attained when the input energy reaches 500 mJ. This characteristic shows promise for this scheme as a future supercontinuum source covering the entire mid-infrared spectral range.

Our work addresses a critical gap in the development of plasma-based mid-infrared sources and significantly broadens their potential application scope. This method enables the realization of plasma-based high-repetition-rate broadband sources, and offers unique time-frequency characteristics that unlock prospects for time-stretched infrared spectroscopy [38]. As there is a corresponding relationship between the frequency and time of pulses, the absorption in the spectrum can be directly manifested as changes in the time-domain waveform. Additionally, this approach maintains the characteristic high peak power and short duration of plasma light sources, which is applicable in fixed-angle broadband laser-driven electron scattering (FABLES) experiments [39]. The broadband light source presented in this work is expected to further extend the momentum transfer range of FABLES, thereby enabling its application to larger molecular systems and dynamic structural analysis of more complex chemical reaction processes.

Currently, all the simulations assume a longitudinally uniform central density. Since the intensity of the deceleration field is related to the plasma density, fine-tuning the density profile allows for optimization of the spectral structure according to requirements. In terms of transverse density distribution, since the dispersion effect is related to the laser radius, adjusting the spot size by changing the channel profile is expected to delay the position where the cascade terminates. This would thereby increase the spectral width under the same input conditions. These aspects merit additional detailed study on this subject.

## Materials and Methods

The simulations were conducted using the fully relativistic particle-in-cell code OSIRIS [44]. In the simulations presented in this article, a cylindrical coordinate system was employed. The simulation box had a length of 25.46 μm in the radial (r) direction, discretized into 200 grid cells, and a length of 89.12 μm in the longitudinal (z) direction, discretized into 3500 grid cells. The time step was set to 0.055 fs, and the total simulation length corresponded to 1018.6 μm. The plasma was uniformly distributed longitudinally, while its transverse profile followed a parabolic density distribution given by $n = n_0(1 + \alpha r^2/r_0^2)$. Here, $n_0 = 1.9 \times 10^{19} \text{cm}^{-3}$, $\alpha = 0.227$, and $r_0 = 5.12$ μm, corresponding to a parabolic plasma channel with a matching radius of 5.12 μm. The laser pulse had an energy of 23 mJ, a wavelength of 0.8 μm, and a pulse length of 10 fs. Its longitudinal profile followed a $\sin^2$ envelope to mitigate longitudinal truncation effects, while the transverse profile was Gaussian. It was focused at the simulation entrance (z = 0) to 5.3 μm, with a normalized vector potential of $a_0 = 1.8$ at the focus. The parametric study results in Figure 5 were conducted with different initial $a_0$ and $\rho_{cen}$ as shown in the figures.

Figure 1 was created using VisualPIC [46] based on simulation results.

## Data availability

The data that support the plots within this paper and other findings of this study are available from the corresponding author upon reasonable request.

## Acknowledgements

The simulation work is supported by the Center of High Performance Computing, Tsinghua University

## Author contributions

T.L.Z. and W.L. proposed the concept. T.L.Z. developed the theoretical model and carried out the simulations. T.L.Z., Y.X.H., J.F.H. and W.L. wrote the paper. J.F.H. and W.L. conceived and supervised the project. All authors discussed extensively the results and commented on the manuscript.

## Conflict of interest

The authors declare no competing interests.